\documentclass[12pt]{iopart}

\usepackage{iopams}  
\usepackage{amssymb}
\usepackage[dvips]{graphicx,color}

\newcommand{\nr}[1]{(\ref{#1})}
\newcommand{\lqcd}{\Lambda_{\mathrm{QCD}}}

\newcommand{\qs}{Q_{\mathrm{s}}}
\newcommand{\ra}{R_A}
\newcommand{\nf}{N_\mathrm{f}}

\newcommand{\as}{\alpha_{\mathrm{s}}}

\newcommand{\xt}{\mathbf{x}_T}
\newcommand{\yt}{\mathbf{y}_T}

\renewcommand{\pt}{{\mathbf{p}_T}}
\newcommand{\qt}{{\mathbf{q}_T}}
\newcommand{\kt}{{\mathbf{k}_T}}

\newcommand{\nabt}{{\boldsymbol{\nabla}_T}}

\newcommand{\gev}{\textrm{ GeV}}

\begin{document}

\title{
Chemical composition of the decaying glasma}

\author{T. Lappi}

\address{
Physics Department, Brookhaven
National Laboratory, Upton, NY 11973, USA}
\ead{tvv@quark.phy.bnl.gov}
\begin{abstract}
The the initial stage of a relativistic heavy ion collision
can be described by a classical color field configuration known as
the Glasma.
The production of quark pairs from this background field is then computed
nonperturbatively by numerically solving the Dirac equation in the classical 
background. The result seems to point towards an early 
chemical equilibration of the plasma.
\end{abstract}



\section{Introduction}

At large energies (small $x$)
the hadron/nucleus wavefunction is characterized by a saturation scale 
$\qs$ arising from the strong nonlinear interactions 
between the wee partons, mostly gluons. If the energy is high enough 
so that $\qs \gg \lqcd$, weak coupling methods can be used to 
decribe physics at transverse momenta $\sim \qs$. In the context of ultrarelativistic heavy ion collisions this means that one can 
hope to understand not only hard probes, but the bulk of particle production 
in terms of weak coupling, deconfined physics. Note that although the 
coupling is weak, the saturated color fields are strong $A_\mu \sim 1/g$,
and the physics must still be treated nonperturbatively.
 
The purpose of this talk is to explore the initial state of a heavy ion collision in this framework. We will first describe what has been called 
the \emph{Glasma} \cite{Lappi:2006fp}, 
a highly coherent classical field in the initial stage
of the collision. We will then move on to study how quark pairs
can be produced from  this, to a first approximation, purely gluonic
system, moving it closer to a chemically equilibrated 
plasma of both quarks and gluons \cite{Gelis:2004jp,Gelis:2005pb}.

\section{Glass and Glasma}

The small $x$ wavefunction of a hadron or nucleus, characterized 
by nonperturbatively large color fields, can be described in terms
of a classical Weizs\"acker-Williams (WW) field radiated by the
hard, large $x$, sources. 
Because of their high speed and Lorentz time dilation, the hard degrees of freedom are seen by the low $x$ fields as slowly evolving in lightcone time. They can therefore be thought of as 
classical, static (in light cone time) sources for the small $x$ 
fields \cite{McLerran:1994ni}.
This effective description has been called
the Color Glass Condensate. The earliest stage of an ultrarelativistic
heavy ion collision is a coherent, classical field configuration of two colliding sheets of Color Glass. This first fraction of a fermi
of the collision, in transition from two sheets of Color Glass into eventually
a quark gluon plasma, is what we refer to as the Glasma \cite{Lappi:2006fp}.

\begin{figure}
\includegraphics[width=0.44\textwidth]{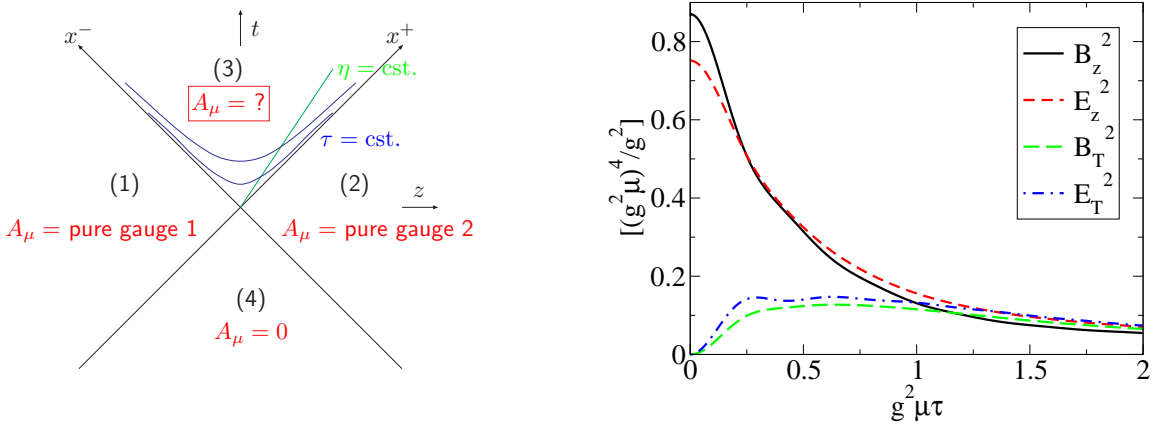}
\hfill
\includegraphics[width=0.46\textwidth]{components.eps}
\caption{Left: The $A_\tau = 0 $-gauge field in different regions of 
spacetime. Right: Components of the glasma field as a function
of $g^2\mu\tau$ in units of $ (g^2\mu)^4/g^2 $
}
\label{fig:glasma}
\end{figure}

\begin{figure}
\includegraphics[width=0.45\textwidth]{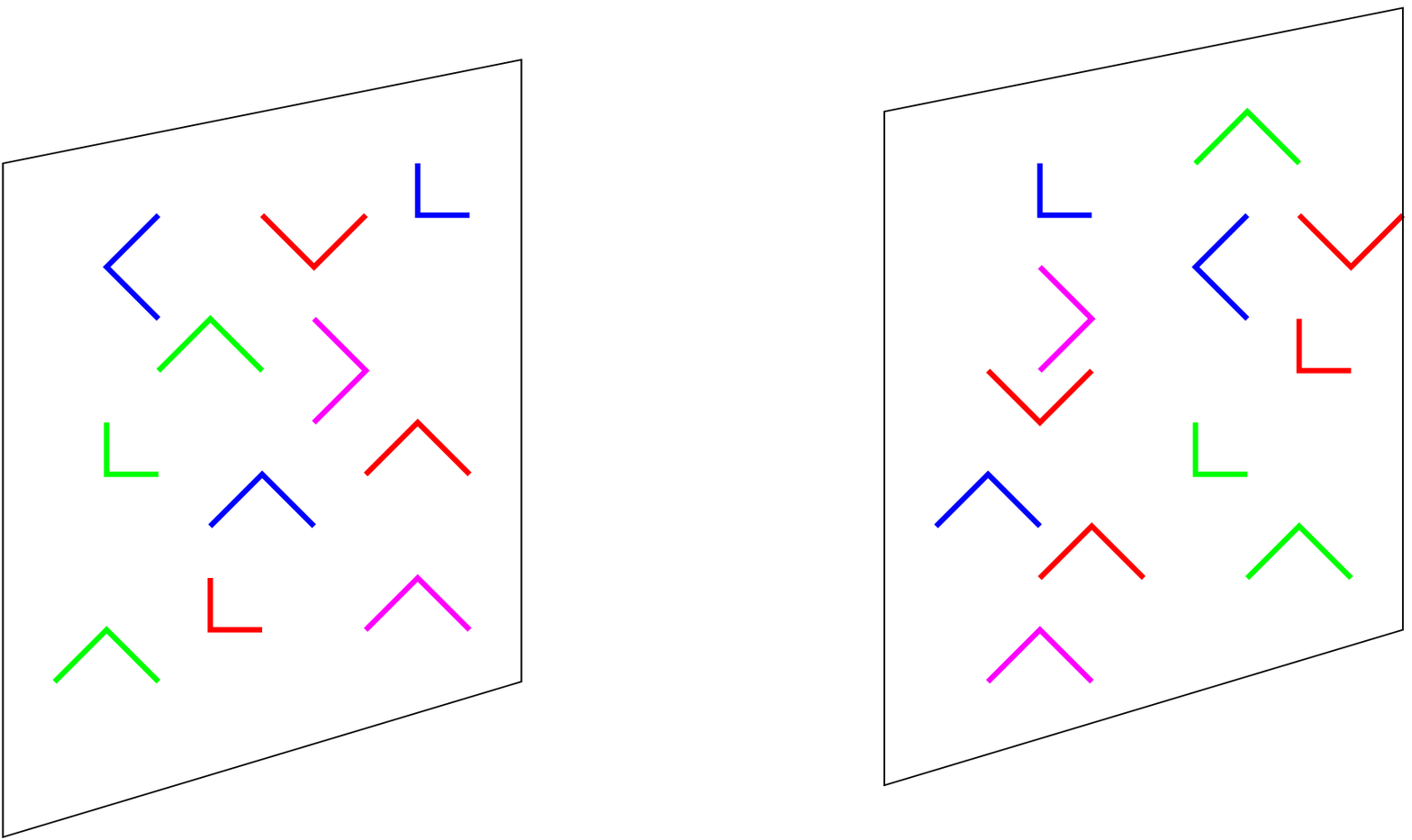}
\hfill
\includegraphics[width=0.45\textwidth]{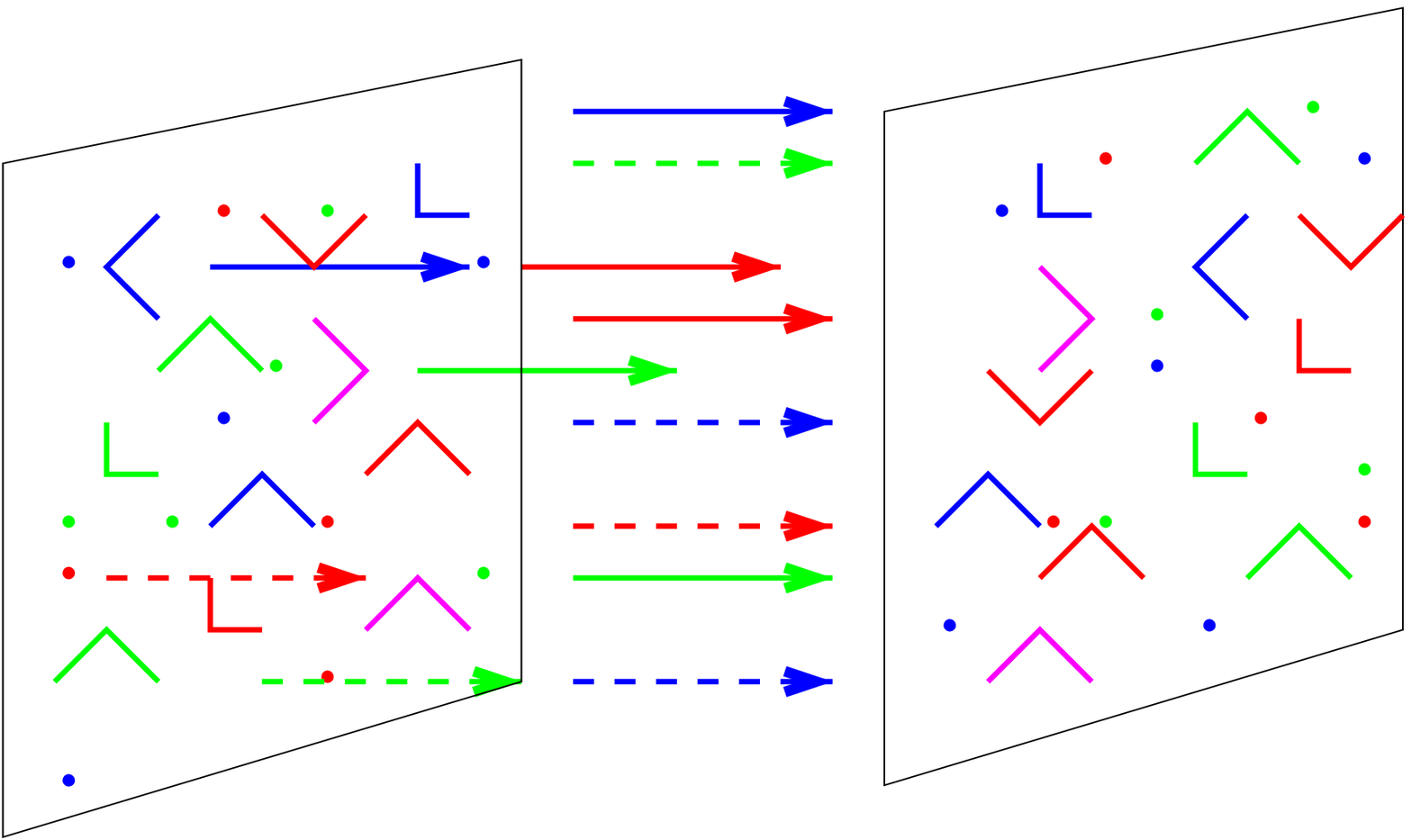}
\caption{The WW fields of the two nuclei before and after
the collision. Before the collision there are only transverse fields
on the sheets. After the collision the interaction of these fields
with the pure gauge field of the other nucleus leaves behind 
an effective electric and magnetic charge density (the dots on the figure) 
on the sheet, and a longitudinal electric and magnetic field between these
effective  charges.}
\label{fig:sheetonsheet}
\end{figure}

The color fields of the two nuclei are transverse electric and magnetic 
fields on the light cone. 
The glasma fields left over in the region between the two nuclei 
after the collision at times $1 \leq \tau \leq 1/\qs$
are, however, longitudinal along the beam axis. One way of 
understanding these field configurations is the following.
Let us work in light cone gauge, so that each nucleus, when going past
a point on the beam axis with no gauge field before the collision, leaves
behind it a pure gauge field (see Fig.~\ref{fig:glasma}).
One can define an effective chromoelectric and chromomagnetic charge density by 
separating the nonlinear parts of the vacuum 
Gauss law and Bianchi identities
\begin{equation}
\left[ D_i,E^i \right] = 0 \quad \textrm{ and } \quad 
\left[ D_i,B^i \right] = 0 
\end{equation}
as
\begin{equation}
\partial_i E^i  = \rho_\mathrm{e} =  i g [A^i,E^i] \quad \textrm{ and } \quad
\partial_i B^i = \rho_\mathrm{m} = i g [A^i,B^i].
\end{equation}
Now we can interpret the interaction of the WW chromoelectric and -magnetic fields of the nucleus on the $x^+$-light cone with the pure gauge field left behind by the other nucleus as an effective chromoelectric
and -magnetic charge density left behind on the light cone. An exactly
opposite charge density is left behind on the other sheet, leading to a longitudinal chromoelectric and -magnetic field between the sheets
\footnote{
Note that the
initial fields being longitudinal in along the beam axis direction
is in no contradiction with the lowest order perturbative description
of the process as $gg \to g$ scattering, because the longitudinal 
(with respect to the beam axis) fields are perpendicular
to the momentum  of the gluon being produced. The initial
polarization state of this gluon is, however, a very particular one.
}. This
structure is illustrated in Fig.~\ref{fig:sheetonsheet}.

As the initial condition has a chromoelectric and -magnetic field 
both in the longitudinal direction, it also has a nonzero Chern-Simons charge
density 
$\sim g^2 \epsilon^{\mu \nu \rho \sigma} \tr F_{\mu \nu} F_{\rho \sigma}$, fluctuating on transverse length scales of the inverse saturation momentum.
This can naturally lead to large P- and CP-violations in these domains, which could have interesting consequences as parity-odd global observables
in the produced mesons 
\cite{Kharzeev:1998kz,Kharzeev:1999cz,Kharzeev:2004ey,Selyuzhenkov:2005xa}.
If the gauge field configurations are exactly boost invariant, 
no topologically nontrivial configurations (sphaleron transitions)
are allowed \cite{Kharzeev:2001ev}, and the effect is suppressed. But
as giving up the assumption of boost invariance in the numerical calculation
is becoming more feasible \cite{Romatschke:2005pm,Romatschke:2006nk}
one will be able to understand this phenomenon better.

How, then, does the glasma decay into plasma, i.e. how can we move from 
a description of the initial state in terms of classical fields into
one formulated for quantum particles? The glasma fields depend on the transverse 
coordinate on a length scale of order $1/\qs$. Thus
to lowest order (in the coupling $\as$ or in $\hbar$) the fields
simply radiate away as gluons with $\pt \sim \qs$. 
As the system expands the fields are diluted and can 
be treated as particles. This lowest order production is the contribution
that is computed in the numerical computations of gluon production
in heavy ion collisions 
\cite{Krasnitz:1998ns,Krasnitz:1999wc,Krasnitz:2000gz,Krasnitz:2001qu,Lappi:2003bi}.

In the rest of this paper we will be concerned with the next order in
$g$ or $\hbar$, namely quantum pair production from the classical background
field. Both gluons and quarks carry color charge and can be produced in 
pairs from the background field. The production of gluon pairs
is the first quantum correction to the classical background field. To
consistently compute it one must take care to avoid double counting
the gluons that are already effectively included in the classical 
background field. A calculation of gluon pair production must be done
consistently with the high energy renormalization group evolution of the 
sources. This computation has not yet been precisely formulated.
On the other hand, computing quark pair production from the classical
background field is more directly relevant for the phenomenology of heavy
ion collisions and easier to formulate, if not necessarily trivial
to carry out nonperturbatively, because it only involves the lowest 
order classical gluon field.

\section{Background field from the MV model}

The hard degrees of freedom are modeled as
classical sources on the light cones:
\begin{equation}
J^{\mu} =\delta^{\mu +}\rho_{(1)}(\xt)\delta(x^-)
+ \delta^{\mu -}\rho_{(2)}(\xt)\delta(x^+).
\end{equation}
The Weizs\"acker-Williams fields describing the softer degrees of freedom can then be computed from the classical Yang-Mills equation
$ [D_{\mu},F^{\mu \nu}] = J^{\nu}$.
In the light cone gauge the field of one nucleus is a pure gauge outside
the light cone (see Fig.~\ref{fig:glasma})
\begin{equation}\label{eq:pureg}
A^i_{(1,2)} = \frac{i}{g} U_{(1,2)} \partial_i U^\dag_{(1,2)},
\textrm{ with }
U_{(m)}(\xt) = \exp \left\{-i g  \frac{\rho_{(m)}(\xt)}{\nabt^2} \right\}.
\end{equation}
In the original MV model, which we shall be using here, the color charge
densities are stochastic Gaussian random variables on the transverse 
plane
\begin{equation}
\langle \rho^a(\xt) \rho^b(\yt) \rangle 
= g^2 \mu^2 \delta^{ab}\delta^2(\xt-\yt),
\end{equation}
where the density of color charges $g^2\mu$ is, up to a numerical 
constant and a logarithmic uncertainty, proportional to the
saturation scale $\qs$.

The initial conditions for the fields in the future light cone between the
two colliding sheets were derived and the equations of motion solved
to lowest order in the fields in Refs.~\cite{Kovner:1995ts,Gyulassy:1997vt} 
(see also Ref.~\cite{Kovchegov:1997ke} for the same calculation in covariant 
gauge and Ref.~\cite{Fries:2006pv} for another formulation of the same lowest order result.) This initial condition has a simple expression in terms
of the pure gauge fields \nr{eq:pureg} of the two colliding nuclei:
\begin{equation}
A^i = A^i_{(1)} + A^i_{(1)} 
\quad \textrm{ and }\quad
A^\eta = \frac{ig}{2} \left[ A^i_{(1)} , A^i_{(2)} \right].
\end{equation}

The solution of the Yang Mills equations beyond leading order is not known 
analytically. They can, however, be solved numerically on the lattice.
The numerical setup was formulated in Ref.~\cite{Krasnitz:1998ns} 
and used to calculate the energy density
and gluon multiplicity corresponding to the glasma fields in 
e.g.  Refs.~\cite{Krasnitz:1998ns,Krasnitz:1999wc,Krasnitz:2000gz,Krasnitz:2001qu,Lappi:2003bi}.
It is these numerically computed color field configurations that we will 
use as the background field for computing quark production 
in the following.

\begin{figure}
\includegraphics[width=0.353\textwidth]{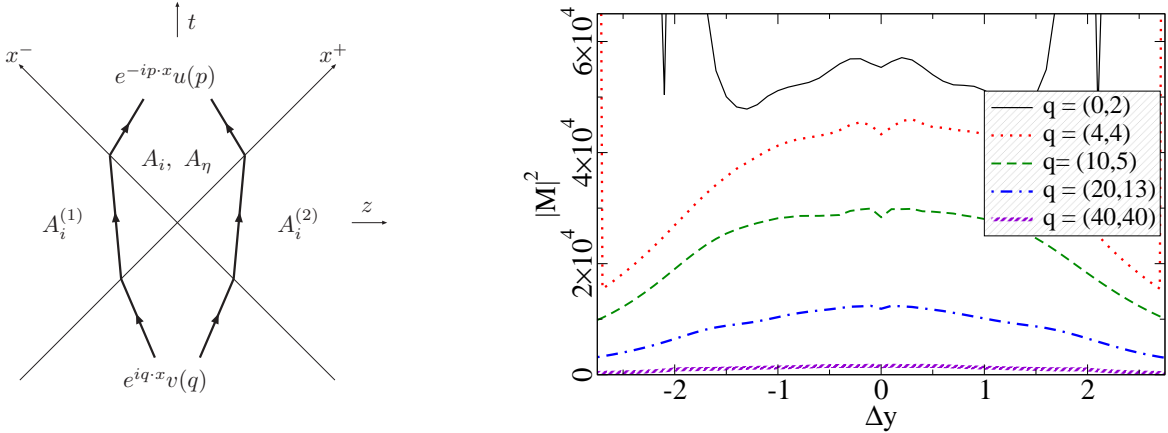}
\hfill
\includegraphics[width=0.547\textwidth]{ampliyp.eps}
\caption{Left: Spacetime in Dirac equation. In the case when the nuclei move 
exactly at the speed of light the solution decomposes into two distinct paths
depending on whether the $x^+$ or $x^-$ light cone is encoutered first.
In the Abelian case these correspond to the $t$ and $u$ channel terms of the
amplitude. 
Right: amplitude as a function of rapidity $y_p - y_q$
for different transverse momentum modes $\qt$.}\label{fig:spacet}\label{fig:ampliy}
\end{figure}

\section{Pair production}

Let us then move on to the computation of
quark pair production from the background
color field described above starting with a few trivial remarks:
\begin{itemize}
\item We are assuming such a high collision energy that 
the nuclear wavefunction is completely gluonic and the process
$gg\to q\bar{q}$ dominates over quark production from the sea and 
valence distributions by processes like $qg\to qg$. This might
not yet be a very good approximation at RHIC energies \cite{Eskola:1996ce}.
We are also neglecting the backreaction of the produced pairs
on the gluon fields.
\item When the computation of gluon production to the lowest
order is done nonperturbatively to all orders in the strong field
the multiplicity and energy density of the gluons is infrared finite.
But for the
computation of quark pair production it is not a priori evident whether 
the limit $m_q\to 0$ is finite.
\item If quark production is dominated by saturation physics,
one could expect production to be flavor blind for $m_q \ll \qs$. Thus 
strangeness would, from the beginning, be equilibrated with 
the light flavors.
\end{itemize}

Our method of computing quark pair production by solving the Dirac 
equation is explained in more detail in Ref.~\cite{Gelis:2004jp}. One 
starts, at $t \to -\infty $ with a negative energy plane wave solution
of the free Dirac equation, $e^{i q \cdot x} v(q)$ (see 
Fig.~\ref{fig:spacet}). The Dirac equation is solved forward in
time until $t \to \infty$, and the solution projected on positive energy plane
wave states $e^{-ip \cdot x}u(p)$. The amplitude is then interpreted as the
amplitude to produce a quark pair from the classical background field.
Solving the Dirac equation forward in time is equivalent to computing
the full retarded (not Feynman) Dirac propagator in the background field.
This means \cite{Baltz:2001dp} that one is computing the expectation
value of the number of pairs produced, not the cross section to produce
exactly one pair (which would be related to the Feynman propagator).
The formal derivation of this procedure is given in Ref.~\cite{Baltz:2001dp},
but the physical interpretation is perhaps more intuitive in terms
of the ``Dirac sea''. Picturing the vacuum as a Dirac sea with all
the negative energy states filled, one is taking a component of the 
``in'' vacuum (at $t \to -\infty $) and computing its overlap with an
``out'' state (at $t \to \infty $) of a quark with momentum $p$. 
If this overlap is nonzero, the background field has lifted a negative 
energy solution from the Dirac sea to positive energy, leaving a hole 
in the sea. This process is then interpreted as the creation of 
a pair of fermions.

Because the background field of one nucleus is a pure gauge
(see Figs.~\ref{fig:glasma} and~\ref{fig:spacet}), the Dirac 
equation can be solved analytically up to the future light cone 
($x^+ =0$, $x^->0$ and $x^- =0$, $x^+>0$). In the Abelian case
also the field inside the future light cone is a pure gauge, and the
whole computation (electron positron pairs
in ultraperipheral collisions of two heavy ions) can be carried through analytically  \cite{Baltz:1998zb}. In our case the field inside the future
light cone is only known numerically, and so we must solve the Dirac equation
numerically starting from an analytically derived initial 
condition at $\tau=0$. It turns out that this is most conveniently
done in the $(\tau,z,\xt)$--coordinate system. The natural choice 
for the temporal coordinate is $\tau$, because the initial condition
for the numerical computation is given at $\tau=0$, and thus no other time
coordinate would lead to a pure initial value problem. For the longitudinal 
coordinate one must (unlike the three dimensional computation of the color field
in Refs.~\cite{Romatschke:2005pm,Romatschke:2006nk}) choose a dimensional coordinate in order to represent the longitudinal momenta ($q^\pm$) on a 
coordinate space lattice at $\tau=0$.

It was shown in Ref.~\cite{Gelis:2003vh} that in the lowest 
this calculation of pair production reduces to a standard 
expression in $\kt$--factorized perturbation theory. The 
``pA'' case, when only one of the classical color fields is treated 
to lowest order, can also be solved analytically
\cite{Fujii:2005vj}.
A $\kt$--factorized perturbative formalism has also been used to 
compute production of heavy quarks in from the 
``Color Glass Condensate'' gluon distributions e.g. in 
Ref.~\cite{Kharzeev:2003sk}. 
A way of approaching the problem from the other limit is presented in
Ref.~\cite{Kharzeev:2006zm}, where, using the WKB approximation, pair 
production is computed nonperturbatively in a short longitudinal color
field pulse, neglecting the magnetic field. Because this computation
also neglects the transverse coordinate dependence of the field, it 
cannot reduce to the same perturbative expression in the weak field limit.

\begin{figure}
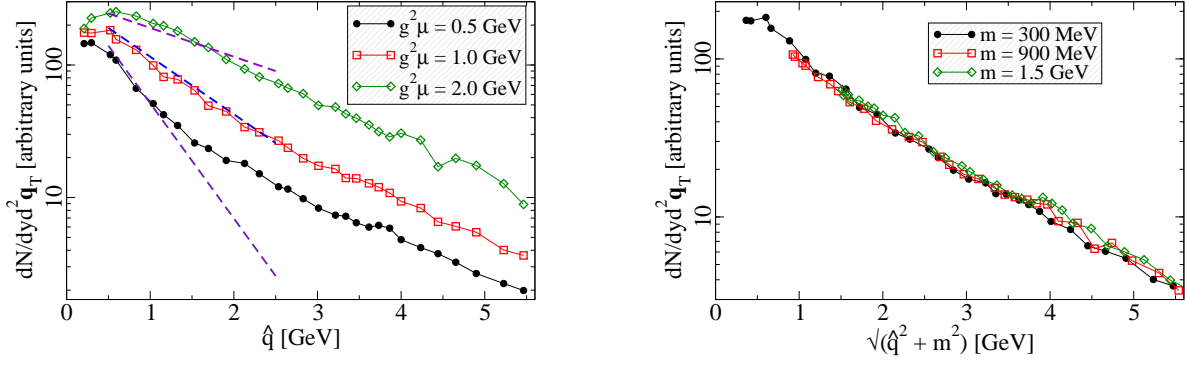

\includegraphics[width=0.45\textwidth]{pspectqslog2p.eps}
\hfill
\includegraphics[width=0.45\textwidth]{pspectlogmtp.eps}
\caption{Left: Antiquark transverse momentum spectrum 
for different values of the classical color charge density
$g^2 \mu$ for $m=0.3 \gev$. The straight lines to (mis)guide the 
eye are $\sim e^{-\qt/(g^2\mu)}$.
Right: The antiquark spectrum as a function of the transverse mass
$\sqrt{\qt^2 + m^2}$ for different masses and $g^2\mu = 1 \gev$.
}\label{fig:spects}
\end{figure}

\begin{figure}
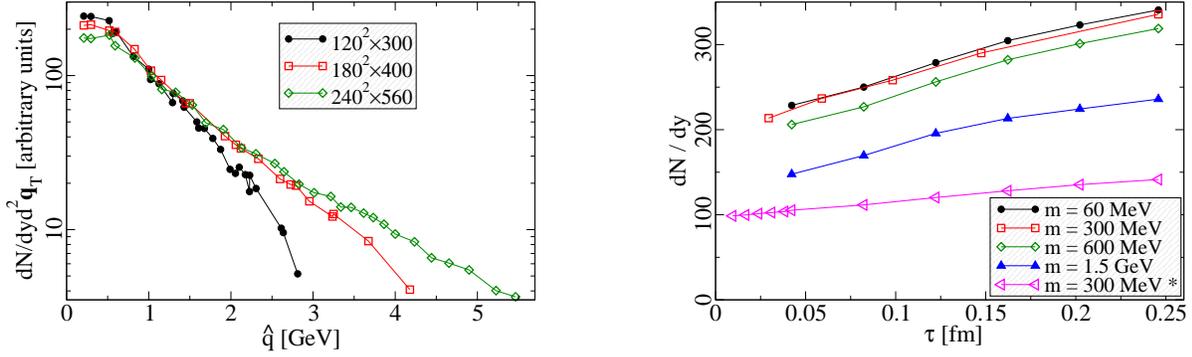

\includegraphics[width=0.45\textwidth]{contlimlogp.eps}
\hfill
\includegraphics[width=0.45\textwidth]{taudep2gevplusbp.eps}
\caption{ Left: Amplitude with $m=0.3\gev$ and $g^2\mu = 1 \gev$
for different lattice sizes, corresponding
to different values of the ultraviolet lattice cutoff.
Right: Amplitude as a function of the time at which the projection
to positive energy states is done. The first four data sets
are for $g^2\mu = 2 \gev$ and the last one (marked with an asterisk) for
$g^2\mu = 1 \gev$.
} \label{fig:contlim} \label{fig:taudep}
\end{figure}

Let us then finally move to the results of the numerical computation
\cite{Gelis:2005pb}. The computation has three independent numerical 
parameters, the color charge density $g^2\mu$ that determines the strength
of the background field, the nuclear radius $\ra$ giving the size of 
the system in the transverse plane, and $m$, the quark mass.
Figure~\ref{fig:ampliy} shows the amplitude,
integrated over the momentum of the quark but for different
transverse momenta of the antiquark, as a function
of the rapidity difference $\Delta y \equiv y_p - y_q$ 
between the quark and the antiquark.
Because the background field is boost invariant, this amplitude only depends
on $\Delta y$ and not the rapidity of the whole system
$(y_p+y_q)/2$. When also integrating over $\Delta y$
one gets the spectrum as a function of
the transverse momentum of the antiquark, shown in Fig.~\ref{fig:spects}
for different strengths of the background color fields.
On a finite lattice the large $\pt$ part of the spectrum is affected
by the proximity of the lattice cutoff $\sim 1/a$, which is 
demonstrated in Fig.~\ref{fig:contlim}, where the same spectrum is plotted
for different values of the lattice cutoff.
Although physically the quarks can only be interpreted as on shell particles
after a formation time $\sim 1/m_\mathrm{T}$, the projection on positive energy
estates can be done at arbitrarily early times. A shown in 
Fig.~\ref{fig:taudep} the amplitude is close to its final value for very
early times close to the light cone.

According to conventional wisdom the initial state of a heavy ion collision is dominated by gluons. Assuming that the subsequent evolution of the system conserves entropy this would mean 
$\sim 1000$ gluons in a unit of rapidity. In the classical field model this corresponds \cite{Lappi:2003bi} to $g^2\mu = 2 \gev$. Our result seems to point to a rather large number of quark pairs present already in the initial state. One could envisage a scenario where, for $g^2\mu = 1.3 \gev$,
these 1000 particles could consist of 400 gluons, 300 quarks and 300 antiquarks (take the lowest curve from Fig.~\ref{fig:taudep} and mutiply by $\nf =3$). 
This would be close to the thermal ratio of $N_q/N_g = 9 \nf/32$.

\section{Conclusions}

We have pointed out some known, but not always fully appreciated, features
of the classical field description of the early stages of a heavy ion 
collision. The ``Glasma'' fields are initially longitudinal chromoelectric
and chromomagnetic fields, of equal magnitude. These fields 
can be thought of as arising from effective chromoelecric and -magnetic
charge densities on the transverse planes caused by the color rotation
of the Weizs\"acker-Williams fields of one nucleus in the pure gauge field
of the other.

We have then studied quark pair production from classical background field of 
McLerran-Venugopalan model studied by solving the 3+1--dimensional Dirac 
equation numerically in this classical background field.
We find that the number of quarks produced large, potentially leading to a very
early chemical equilibration of the quark and gluon degrees of freedom.
But in order to put this conclusion on a more theoretically 
sound footing one must also compute gluon production to the same order in
the couopling $\as$, which has not yet been done. Our numerical method
does not yet allow us to extend our method to large quark masses or
transverse momenta.

\section*{Acknowledgments}

The work presented here has been done in collaboration with F. Gelis, K. Kajantie and  L. McLerran.
 The author wishes to thank R. Venugopalan, K. Tuchin and D. Kharzeev for many discussions on the subject. 
This research has been supported by the U. S. Department of Energy
under Contract No. DE-AC02-98CH10886. 

\section*{References}

\bibliographystyle{JHEP-2mod}
\bibliography{spires}

\end{document}